\documentclass[12pt,preprint]{aastex}

\shorttitle{NLTE Oxygen and Iron abundances in HD140283}
\shortauthors{Shchukina \& Trujillo Bueno}

\begin{document}

\title{The impact of non-LTE effects and granulation inhomogeneities
on the derived iron and oxygen abundances
in metal-poor halo stars}

\author{N.G. Shchukina}
\affil{Main Astronomical Observatory, National Academy of
Sciences, Zabolotnogo st., 27, \\03680 Kyiv, Ukraine}
\email{shchukin@mao.kiev.ua}

\author{J. Trujillo Bueno\altaffilmark{1}}
\affil{Instituto de Astrof\'{\i}sica de Canarias, 38205 La Laguna,
Tenerife, Spain} \email{jtb@iac.es}

\and

\author{M. Asplund}
\affil{Research School of Astronomy and Astrophysics, Mt Stromlo
Observatory, Cotter Road, Weston, ACT 2611, Australia}
\email{martin@mso.anu.edu.au}

\altaffiltext{1}{Consejo Superior de Investigaciones
Cient\'{\i}ficas; Spain}


\begin{abstract}

This paper presents the results of a detailed theoretical
investigation of the impact of non-LTE effects and of granulation
inhomogeneities on the derived iron and oxygen abundances in the
metal-poor halo subgiant HD140283. 
Our analysis is based on both
the `classical' one-dimensional (1D) stellar atmosphere models and
on the new generation of three-dimensional (3D) hydrodynamical
models. The non-LTE calculations presented here
have been carried out without
inelastic collisions with neutral hydrogen atoms.
We find that if non-LTE effects are taken into
account when synthetizing the Fe\, {\sc i} spectrum in both type
of models, then the derived iron abundance turns out to be very
similar in both cases. The emergent spectral line profiles in both models
are very much weaker in non-LTE than in LTE because the
UV overionization mechanism produces a very strong underpopulation
of the Fe\, {\sc i} levels, in particular in the granular regions of the 3D model. 
As a result, the non-LTE effects on the derived iron abundance are very
important, amounting to ${\sim}0.9$ dex and to ${\sim}0.6$ dex
in the 3D and 1D cases, respectively. On the other hand, we find that
non-LTE and 3D effects have to be taken into account for a
reliable determination of the iron abundance from weak
Fe\, {\sc ii} lines, because the significant overexcitation of
their upper levels in the granular regions tend to produce
emission features. As a result such Fe\, {\sc ii} lines are weaker
than in LTE and the abundance correction amounts to ${\sim}0.4$
dex for the 3D case.

We derive also the oxygen-to-iron abundance ratio in the
metal-poor star HD140283  by using the O\,{\sc i} triplet at
7772--5 \AA\     and the forbidden [O\,{\sc i}] line at 6300 \AA. Our
3D results for the oxygen abundance confirm the values reported in
some recent investigations. While the oxygen abundance derived
from the O\,{\sc i} IR triplet is not very sensitive to the
presence of granulation inhomogeneities, such 3D effects amount to
${\sim}-0.2$~dex for the [O\,{\sc i}] line. The non-LTE abundance
correction for the O\,{\sc i} IR triplet turns out to be
$-0.2$~dex, approximately.

Interestingly, when both non-LTE and 3D effects are taken
into account there still remains significant discrepancies in the
iron abundances derived from Fe {\sc i} and Fe {\sc ii} lines, as
well as in the oxygen abundances inferred from the O\,{\sc i} and
[O\,{\sc i}] lines. We conclude that the discrepancies could be due to
uncertainties in the stellar parameters of this metal poor star.
We argue that adopting $T_{\rm
eff}{\approx}5600$ K
(instead of $T_{\rm eff}{\approx}5700$ K) and
[Fe/H]${\approx}-2.0$ (instead of [Fe/H]${\approx}-2.5$)
substantially reduces the discrepancies in the abundances of iron
and oxygen inferred from several spectral lines. Under such
circumstances we find [O/Fe]${\approx}0.5$ at [Fe/H]=$-2$.
Obviously, our tentative
conclusion that the metalicity of this type of metal-poor stars is
significantly larger than previously thought may have far-reaching
implications in stellar astrophysics.

\end{abstract}



\section{Introduction}


%
Reliable determinations of the oxygen and iron abundances in the
atmospheres of metal-poor stars are of paramount importance for
21st Century astrophysics. This is because the abundances of
oxygen and iron in metal-poor stars represent fundamental
astrophysical quantities for determining the ages of the oldest
stars, for constraining the models of the chemical evolution of
the Galaxy, for estimating the yields of light elements (lithium,
beryllium and boron) from spallation processes in the interstellar
and circumstellar gas, for testing different scenarios of
nucleosynthesis in supernovae, etc. Such studies largely rely upon
the oxygen-to-iron abundance ratio
(${\rm [O/Fe]}={\rm log}{\rm (O/Fe)}_{\star}-{\rm log}{\rm
(O/Fe)}_{\odot}$) versus the stellar metallicity [Fe/H].


\subsection{The oxygen abundance problem in metal-poor stars}

The shape and spread of the [O/Fe] vs. [Fe/H] relation is still
hotly debated in the literature (e.g.,
%
%
Israelian et al. 2001;
Asplund \& Garc\'\i a P\'erez 2001;
Mel\'{e}ndez \& Barbuy 2002;
Nissen et al. 2002;
Fulbright \& Johnson 2003;
Takeda 2003,
and references therein).





%


%




%


The controversy is linked with discordant results
obtained using different indicators of the oxygen abundance, such
as:
i) the  oxygen  [O\,{\sc i}] forbidden  $\lambda$6300 \AA\
line;
ii) the oxygen O\,{\sc i} infrared triplet at 7772--5 \AA;
iii) the OH lines in the near ultraviolet;
iv) the OH lines in the infrared.
More precisely, the oxygen abundance derived using all these lines
agree
more or less
with each other for moderate metal-deficient disk stars
having $-1 \le {\rm [Fe/H]} \le 0$.
However, there are clear discrepancies among the results obtained
for metal-poor halo stars with lower metallicities.
%
%
Some authors find that the [O/Fe] ratio
%
shows a plateau  between 0.4 and 0.5 dex in the metallicity
range -3${<}$[Fe/H]${<}-1$
%
(Barbuy 1988;
Kraft et al. 1992;
Fulbright \& Kraft 1999;
Asplund \& Garc\'{\i}a P\'{e}rez 2001;
Caretta et al. 2000;
Mel\'{e}ndez \& Barbuy 2002),
while others find a linear increase
with decreasing metallicity, reaching
[O/Fe]$\simeq +1$~dex at [Fe/H]$=-3$
%
(Abia \& Rebolo 1989;
Tomkin et al. 1992;
Cavallo et al. 1997;
Israelian et al. 1998, 2001;
Boesgaard et al. 1999;
%
%
Mishenina et al. 2000;
and references therein).
%
%
%
%
%
Recently, Nissen et al. (2002) have
analyzed [O\,{\sc i}] and [Fe\,{\sc ii}] lines within the
framework of 3D atmospheric models and LTE line formation
concluding that there is a quasi-linear trend for [Fe/H]$<-2$,
with [O/Fe]$\simeq +0.5$~dex at [Fe/H]$=-2.5$.
Furthermore, an extensive 1D re-analysis of published equivalent
widths for the {O\,{\sc i}} IR and the forbidden [{O\,{\sc i}}]
lines
(see Takeda 2003)
shows that the discrepancy between O\,{\sc i} and [O\,{\sc i}]
lines in metal-poor halo stars tends to be smaller for stars with
higher effective temperatures ($T_{\rm eff}$) and gravities (${\rm
log} g$), i.e. preferentially for dwarfs and subgiants than for
giants.
Fulbright \& Johnson (2003) support this conclusion.
In addition, the oxygen abundances in dwarfs derived from the
{O\,{\sc i}} IR lines tend to have values between those obtained
with the UV OH and [O\,{\sc i}] lines.


\subsection{The iron abundance problem in metal-poor stars}

A key problem directly related with the [O/Fe] issue in
metal-poor stars is the reliability of the iron abundance
determinations. In fact, the derivation of the iron abundance from
Fe\,{\sc i} and/or Fe\,{\sc ii} lines is a very complex problem
that requires detailed non-LTE investigations (hereafter NLTE)
like the one carried by Shchukina \& Trujillo Bueno
(2001) and in this paper. In general, the Fe\,{\sc i} lines suffer
from UV overionization in the atmospheres of late-type stars
%
%
(Athay and Lites 1972; Rutten 1988; Shchukina \& Trujillo Bueno
2001).
This NLTE effect tends to be substantially more important in
metal-poor stars than in solar-like stars as a result of the lower
electron density and of the weaker UV blanketing
(Th\'{e}venin \& Idiart 1999; Shchukina and Trujillo Bueno 2001;
Israelian et al. 2001, 2004;
Shchukina et al. 2003). We should mention also that
Gratton et al.(1999)
concluded that NLTE effects are rather small in metal-poor stars.
However, their calculations presumably overestimated the
efficiency of inelastic collisions with hydrogen
atoms which thermalize the Fe\,{\sc i} levels and thus almost
restore LTE conditions.

It is of interest to point out that while the Fe\,{\sc i} lines
are (in principle) expected to be sensitive to the UV
overionization mechanism, the LTE approximation is thought to be
suitable for weak lines of Fe\,{\sc ii} (see, e.g., Asplund et al
1999; Nissen et al. 2002). However, as we shall see below this
does not have to be necessarily the case because many subordinate
Fe\,{\sc ii} lines whose upper levels are of odd parity (starting
at $z\rm{{^6}D^{o}}$) might be partially filled by emission due to
an optical pumping mechanism similar to that investigated by Cram
et al. (1980) for the solar case.
%
%
%


\subsection{The solar oxygen abundance problem}

Obviously, the solar case constitutes a fundamental reference for
a reliable quantification of chemical abundances in other stars.
The solar iron abundance appears to be fairly well determined now
(${\rm log} \epsilon_{\odot}$(Fe)=7.50$\pm 0.10$; see Shchukina \&
Trujillo Bueno 2001) --- that is, it is similar to the meteoritic
value. In fact, this is the value adopted by Nissen et al. (2002)
in their study of the O/Fe ratio in metal-poor stars.

Great progress has been also made towards a reliable
determination of the solar oxygen abundance. The 1D modeling
approach leads to discrepancies between the solar oxygen abundance
determined from the permitted O\,{\sc i} lines and from the
[O\,{\sc i}] and OH lines. Abundance determinations from OH lines
using the 1D model of Holweger \& M$\ddot{\rm u}$ller (1974)
suggested first a high oxygen abundance value (${\rm log}
\epsilon_{\odot}$(O)=8.93$\pm 0.03$; see
Grevesse et al. 1984;
Sauval et al. 1984;
Anders \& Grevesse 1989).
Several years later, Grevesse \& Sauval (1998) obtained an oxygen
abundance that was lower by 0.1 dex. This lower value was found
after using improved ${\rm log} gf$ values and a slightly modified
1D model, which helped to remove the existence of a dependency of
the derived oxygen abundance with the excitation potential of the
lower level of the OH transitions.
The forbidden [O\,{\sc i}] line gives also a high oxygen abundance
value when the contribution of the blend due to a Ni\,{\sc i} line
is ignored
(Lambert, 1978;
Sauval et al. 1984;
Holweger 2001).  However, the permitted O\,{\sc i} lines at
$\lambda$7772--5 \AA\ yield a value for the oxygen abundance
that is lower by 0.2 dex
(Kiselman 1993; Holweger 2001).
Some authors have argued that the existing NLTE calculations for
these spectroscopic indicators of the oxygen abundance might not
be valid
due to the neglect of inelastic collisions with hydrogen. For this
reason the solar oxygen abundance derived  from the [O\,{\sc i}]
forbidden line was considered as the most reliable one, because of
the suitability of the LTE approximation (hence,
uncertainties of collisional rates with hydrogen atoms do not play
any role). Another argument used to support a
high value for the oxygen abundance is the small scatter in the
oxygen abundances obtained from different OH lines.
Kiselman \& Nordlund (1995)
tried to clarify the solar oxygen abundance issue by using a 3D
solar photospheric model that is similar but less sophisticated than that of
Asplund et al. (2000). They concluded that a downward revision
of the solar oxygen abundance is required, with respect to the
value derived from [O\,{\sc i}] and OH lines.
More recently, Allende Prieto et al. (2001) have obtained a more
precise estimate by taking into account the influence of the
above-mentioned blend by the Ni\,{\sc i} line and by using
the  3D hydrodynamical model of Asplund's et al. (2000).
They found
${\rm log}{\epsilon}_{\odot}$(O)=8.69$\pm 0.05$, which shows that
the [O\,{\sc i}]-based abundance derived via the 3D approach is
significantly lower than that derived from classical 1D model
atmospheres.

In a recent IAU Symposium on modeling of stellar atmospheres
Shchukina et al. (2003) reported that the
oxygen abundance obtained via NLTE synthesis of the
solar O\,{\sc i} IR triplet in the
3D solar photosphere model of Asplund's et al. (2000) is
${\rm log}\epsilon_{\odot}$(O)${\approx}8.7$. This result 
is in good agreement with
both Allende Prieto's et al. (2001) conclusion and with the recent
3D analysis of Asplund et al. (2004) who found excellent agreement
between the averaged oxygen abundance values derived from
[O\,{\sc i}], O\,{\sc i}, OH vibration-rotation and OH pure rotation 
lines\footnote{It is interesting to mention that the
oxygen abundances obtained by
Asplund et al. (2004)
from OH rotational lines using the 3D modeling approach show a
large line-to-line scatter, and with sensitivity to the line strength
and to the excitation potential of the transition's lower level.
However, such presumably minor discrepancies
are not the topic of this paper. Here
we restrict ourselves to consider the NLTE determination of the
solar oxygen abundance via
O\,{\sc i} lines, for consistency with similar computations
we have carried out for the metal-poor star HD140283.}.


\subsection{Possible reasons for the abundance discrepancies}

It is currently believed that abundance discrepancies in
metal-poor halo stars and the sun are dominated by deficiencies in
the spectral line modeling. Errors in the measurements of
equivalent widths are of minor importance with the exception of
weak [O\,{\sc i}] and Fe\,{\sc ii} lines.
%
%
%
All abundance indicators in metal-poor  stars are to a different
degree sensitive to the adopted stellar parameters ($T_{\rm eff}$,
${\rm log} g$ and [Fe/H]). They have to  depend also on
atmospheric properties such as the temperature gradient
and the ratio of the line to continuous opacity. Large
uncertainties could result from neglecting atmospheric
inhomogeneities and velocity fields (i.e., granulation
inhomogeneities produced by the stellar surface convection
process). Part of the discrepancies are probably due to NLTE
effects.
There is now an abundance of literature concerning the iron and
oxygen line formation problems in classical plane-parallel stellar
atmospheric models
(Th\'{e}venin \& Idiart 1999;
Mishenina et al. 2000;
Kiselman 2001;
Takeda 2003;
and references therein).
The recent generation of 3D hydrodynamical
models of stellar atmospheres have activated the present debates
on the influence of granulation inhomogeneities and of NLTE
effects on the determination of chemical abundances in metal-poor
stars
%
%
(see
Asplund et al. 1999;
Asplund and Garc\'{\i}a P\'{e}rez et al. 2001;
Shchukina \& Trujillo Bueno 2001;
Nissen et al. 2002;
%
%
%
Asplund et al. 2004). In this respect, it is very important to
emphasize that the 3D hydrodynamical simulations of Asplund and
coworkers show that the atmospheres of metal-poor halo stars may
have substantially lower kinetic temperatures than the radiative
equilibrium temperatures of the `classical' theoretical 1D model atmospheres
of Kurucz (1993) and the MARCS collaboration (Gustafsson et al. 1975).

\subsection{Motivation}

The previous discussion makes evident that detailed, 3D+NLTE
investigations of the iron and oxygen abundances are urgently
needed. However, it is crucial that this type of analyses are
carried out {\em jointly} for iron and oxygen, and using the
same set of atmospheric models,
continuum opacity package and radiative
transfer code.
The present paper presents the results of such an investigation.
We study the NLTE formation problem
of the Fe\,{\sc i}, Fe\,{\sc ii} and O\,{\sc i} lines in a 3D
hydrodynamical model of the atmosphere of the
metal-poor star HD140283.
This classical halo subgiant is widely used for studies of the
very early Galaxy and the origin of the chemical elements. In
order to quantify the impact of 3D simulations on the abundance
determinations in HD140283 we carry out a detailed comparison with
the 1D classical modeling.
To this end, we use the 3D hydrodynamical atmospheric model of
Asplund's et al. (1999) for the star HD140283, as well as a suitable grid of Kurucz's 1D models.
Moreover, we evaluate  the extent to which 3D+NLTE effects
influence the determination of the stellar parameters of this
metal-poor subgiant.

Our paper considers also the case of the Sun as a reference for
stellar abundance determinations with emphasis on the solar oxygen
abundance derived from the O\,{\sc i} IR  triplet. As in our
NLTE+3D analysis of the solar iron abundance
(Shchukina \& Trujillo Bueno 2001)
we investigate here the solar oxygen abundance problem by using
the same 3D hydrodynamical solar granulation model.
%
%


\section{The diagnostic method} \label{methods}

In order to determine the stellar abundances of iron and oxygen we
have to compare some of the properties of the observed spectral
line profiles (e.g., their equivalent widths) with those
calculated via the numerical solution of the radiative transfer
equation in a given model of the atmosphere of the star under
investigation. To this end, we have used the same NLTE code and
strategy that we applied in our previous investigation of the iron
line formation problem in a 3D hydrodynamic model
of the solar photosphere (see Shchukina \& Trujillo Bueno 2001),
with the only difference that we have now computed flux profiles.
Our 3D strategy for investigating the issue of the
abundances of iron and oxygen
in the metal-poor star HD140283 is similar to
that of described by Shchukina \& Trujillo Bueno (2001) for
solar-like atmospheres, but using a 3D snapshot model
from Asplund's et al. (1999) stellar surface convection simulation.
Therefore,
we have neglected the effects of horizontal radiative
transfer on the atomic level populations -that is, we
have used the so-called 1.5D approximation.
As discussed by
Shchukina \& Trujillo Bueno (2001) this is expected to be a good
approximation for determining the stellar iron abundance from
Fe\,{\sc i} and Fe\,{\sc ii} lines.
Kiselman \& Nordlund (1995) already demonstrated that it is also a
good approximation for the oxygen IR triplet. 

Our NLTE code is based on very efficient iterative methods
(Trujillo Bueno \& Fabiani Bendicho 1995; Socas-Navarro \&
Trujillo Bueno 1997; Trujillo Bueno 2003) that were developed to
facilitate radiative transfer simulations with complex atomic
systems in realistic stellar atmospheric models. We have carried
out both LTE and NLTE spectral line syntheses by using both the 1D
and 3D models mentioned below. In each case, the abundances of
iron and oxygen are modified iteratively till achieving the best
fit to the observations.


\section{The observations} \label{observations}

%
We have chosen 33 Fe\,{\sc i} and 15 Fe\,{\sc ii} lines for our 3D
and 1D determinations of the iron abundance in the halo subgiant
HD140283. The equivalent widths ($W$)
of all the Fe\,{\sc i} lines and of 11 Fe\,{\sc ii} lines
were obtained from
high-resolution spectra observed a few years ago with the Harlan
J. Smith 2.7m telescope at McDonald observatory
(R. Garc\'\i a L\'opez, private communication; see also
Allende Prieto et al. 1999).
The equivalent widths ($W$) for the remaining 4 Fe\,{\sc ii}
lines are taken from Table 3 of
Nissen et al. (2002).
%
%
%
They are weak lines
with $W$ in the range
1$-$4 m\AA\ . Their
wavelengths are ${\lambda}{\lambda} 6149.238, 6247.562, 6432.654,
6456.376$ \AA. The iron lines used in our analysis are
listed in Table 1.


The O\,{\sc i} IR triplet lines in the metal-poor star HD140283
are weak.
Given uncertainties in the measurements
we have averaged their equivalent widths over three sets of
observations (Abia \& Rebolo 1989;
%
Tomkin et al. 1992,
%
and
G. Israelian, private communication).
The resulting equivalent widths of the 7772, 7774 and 7775 \AA\
lines used in our oxygen abundance determinations are 7.9,  4.8,
3.4 m\AA, respectively.
The equivalent width of the forbidden  [O\,{\sc i}] $\lambda$6300
\AA\ line ($W=0.5$~m\AA) has been taken from
Nissen's et al. (2002) paper. The oxygen line data are specified
in Table 1.


Measurements of the equivalent widths of the solar O\,{\sc i} IR
lines tend to have an uncertaintly  of ${\sim}$5\%. Therefore, we
have used averaged values of five sets of observations at the
solar disk centre
(Jungfraujoch atlas of Delbouille et al. 1973;
Gurtovenko \& Kostik 1989;
Altrock 1968;
M$\ddot{\rm u}$ller et al. 1968;
King \& Boesgaard 1995).
The resulting averaged solar equivalent widths for the 7771.96, 
7774.18 and 7775.40 \AA\ lines are 81.6, 70.4, 57.2 {m\AA,
respectively.


\section{The atmospheric models} \label{atmos}


In each case (i.e., the solar and the metal-poor star)
we have used a single
3D snapshot model taken from realistic radiation
hydrodynamical simulations of stellar surface convection (Stein \&
Nordlund 1998; Asplund et al. 1999, 2000). Such time-dependent
simulations have no adjustable free parameters, since they are
based on the numerical solution of the equations of mass, momentum
and energy conservation for given values of the effective
temperature ($T_{\rm eff}$), surface gravitational acceleration
(${\rm log}g$) and chemical composition. In order to facilitate
the radiative transfer calculations the original snapshots, which
have 82 gridpoints along the vertical direction, were interpolated
to grids of $50\times50\times102$ and $50\times50\times121$ points
for the solar and metal-poor star cases, respectively. The
improvement in the vertical resolution (i.e., with 102 and 121
gridpoints sampling the line and continuum forming regions) was
introduced to enhance the accuracy of the NLTE calculations.

The solar 3D model is that described by Asplund et al. (2000).
This 3D model of the solar photosphere has been used by Shchukina
\& Trujillo Bueno (2001) in their NLTE investigation of the solar
iron abundance.
We have also used it to determine the solar oxygen abundance via
NLTE analysis of the O\,{\sc i} IR triplet at 7772--5 \AA.

Information on the corresponding 3D model for the star HD140283
can be found in Asplund et al. (1999). 
The adopted stellar parameters were ${\rm
log}g=3.7$, $T_{\rm eff}=5700$ K and [Fe/H]=-2.5, which result
from the infrared flux method (IRFM) for $T_{\rm eff}$, from {\it
Hipparcos} parallaxes for ${\rm log} g$ and from published values
for the [Fe/H] (see Asplund et al. 1999).
%
%
In our differential comparison between the results obtained via
the 1D and 3D approaches we have used atmospheric models
characterized by the
same stellar parameters ($T_{\rm eff}$, ${\rm log}g$ and [Fe/H]).
In Sect.~\ref{discussion}
we will also show results in a grid of `classical' 1D models
spanning the range
$5300 \le T_{\rm eff} \le 5700$,
 $3.7 \le {\rm log}g \le 3.9$,
  $-2.0 \le {\rm [Fe/H]}=-2.5$.


\section{The atomic models} \label{atom}

In our spectral synthesis of oxygen and iron lines we adopted the
quantum-mechanical approach developed by
Anstee \& O'Mara (1995);
Barklem \& O'Mara (1997), and
Barklem et al (1998) for the Van der Waals collisional broadening.
In the following subsections we provide some information about the
atomic models of oxygen and iron we have used for our radiative
transfer calculations.


\subsection{Oxygen} \label{oxygen}

Our atomic model for oxygen is based on the data of
Carlsson \& Judge (1993).
It has 23 O\,{\sc i} fine structure levels and one O\,{\sc ii}
level. We considered 31 bound-bound and 23 bound-free radiative
transitions.
The O\,{\sc i} Grotrian diagram has a helium-like structure. The
triplets and the quintets are only weakly coupled to each other.
The two systems differ in one important aspect: the triplets are
connected to the ground state by strong resonance lines, while the
quintets are not. The latter form a separate system with the
$3{^5}S$ level acting as a ground level and the 7772--5 \AA\ IR
triplet as resonance lines. O\,{\sc i} levels have hydrogen-like
energies. As a result, there is an important wavelength
coincidence of the O\,{\sc i} 1025\,\AA\ line with the hydrogen
${\rm Lyman}-{\beta}$ line.
We emphasize that the NLTE results for O\,{\sc i} are quite
insensitive to including additional levels and transitions
(Shchukina 1987; Kiselman 1993; Takeda 2003).}
In agreement with previously published results we have found that
the effects of the coupling between the triplet and quintet
systems, ${\rm Ly}_{\beta}$ pumping, and the binding of CO
molecules are marginal in the stellar atmospheres studied here
(see
                 Shchukina 1987;
                 Kiselman  1993;
                 Carlsson \& Judge 1993;
                 Kiselman \& Nordlund 1995).


\subsection{Iron} \label{iron}

Our model atom for Fe\,{\sc i}$+$Fe\,{\sc ii}$+$Fe\,{\sc iii} is
similar to that used by Shchukina \& Trujillo Bueno (2001). It has
225 Fe\,{\sc i} levels, 23 Fe\,{\sc ii} levels and 1 level for
Fe\,{\sc iii} including their multiplet fine structure. The
Fe\,{\sc i} levels are interconnected by 330 bound-bound radiative
transitions, while we have 25 strong radiative
transitions among the Fe\,{\sc ii} levels. All levels are
coupled via collisional transitions with electrons. The Fe\,{\sc i} term
diagram is, in fact, complete up to
an excitation potential $\chi=5.72$ eV. At higher
energies
it contains about 50\% of the terms that have been presently
identified. Each term of our atomic model is coupled to its
parent term of the next ionization stage by photoionization
transitions. For most of the  Fe\,{\sc i} terms that have ${\chi}\le$5 eV
we used the experimental photoionization cross sections given by
Athay \& Lites (1972). For the rest of them, and also for the
Fe\,{\sc ii} terms, the hydrogen-like approximation was used.
\footnote{It is interesting to mention that the new radiative cross-sections
of Bautista (1997) are substantially larger than those given by the
hydrogenic approximation. Therefore, in what respects
the influence of photoionization cross sections, we might
perhaps be underestimating the NLTE effects.}

In our calculations we have not included inelastic
collisions with hydrogen atoms. We have also neglected the UV haze
opacities. We have carried out several numerical experiments in
order to investigate how large could be the uncertainties in the iron
abundance determinations due to uncertainties in the hydrogen
collisional rates, in the UV haze opacity and
in the bound-free cross sections.
Considering all factors together we have found that the
uncertaintly for the solar iron abundance does not exceed 0.07
dex (see Shchukina \& Trujillo Bueno 2001). The problem of the iron abundance uncertainty for the metal-poor star HD140283 is discussed in Section 8.


\section{The oxygen abundances} \label{o_sun}

In this section we report on the results we have obtained for the
oxygen abundance in the sun and in the metal-poor star HD140283.
Our abundance determinations and some recent abundance
studies by other authors are briefly summarized in
Tables 2 and 3.

\subsection{The Sun} \label{sun}

We have derived the abundance of oxygen in the solar photosphere
by fitting the observed disk centre profiles and equivalent widths
of the O\,{\sc i} IR triplet via spectral synthesis in the 3D
model of the solar photosphere. Taking into account NLTE effects
we find ${\rm log}\epsilon(\rm O)=8.70 \pm 0.06$ (see also
Shchukina et al. 2003), while the LTE approximation gives ${\rm
log}\epsilon(\rm O)=8.93 \pm 0.06$.
Note, that the uncertainties reflect the line-to-line
scatter for the OI IR triplet lines.
In both cases we obtained an almost perfect fit to the
observed disk centre profiles and equivalent widths. Our NLTE
result is in good agreement with that obtained by
Allende Prieto et al. (2001)
from the forbidden  [O\,{\sc i}] 6300 \AA\  line %
(${\rm log}\epsilon(\rm O)=8.69$).
It is also consistent with the 3D result of Asplund at al. (2004)
for the OH lines (${\rm log}\epsilon(\rm O)=8.61 - 8.65$) and 
with their full 3D+NLTE result for
the O\,{\sc i} triplet (${\rm log}\epsilon(\rm O)=8.65$).
It is important to note that the small discrepancy between ours
and Asplund's et al. (2004) solar abundance value derived from the
O\,{\sc i} IR triplet is mainly due to differences in the oscillator
strengths used ($\Delta {\rm log}gf \approx -0.05$~dex).
The  oxygen abundance changes that result from a 5\%
uncertainty in the solar equivalent width measurements of these
spectral lines lie in the same range ($\approx \pm 0.05$ dex).

The NLTE formation of the O\,{\sc i} triplet  can be easily
understood via a two-level atom approach if the line opacity is
calculated assuming
LTE
(Shchukina 1987;
Kiselman 1993;
Kiselman \& Nordlund 1995;
Takeda 2003;
Shchukina et al. 2003,
Asplund et al. 2004,
and references therein).
The photon losses mechanism lowers the line source functions both
in the granular and intergranular regions of the 3D model, which
implies that the NLTE profiles of the IR triplet lines are deeper
than in LTE. The effect is larger in the intergranular regions,
but the spatially averaged emergent profiles are only slightly
affected by the granulation inhomogeneities. Consequently, the
effects of granulation inhomogeneities
on the NLTE solar oxygen abundance determination are
small, i.e. $\sim 0.06$ dex
(Shchukina et al. 2003).
More significant are the NLTE
abundance corrections for the O\,{\sc i} infrared triplet lines
observed at the solar disk centre, which amount to $-0.25$,
$-0.22$, $-0.20$ dex for
${\lambda}{\lambda} 7771.96, 7774.18, 7775.40$ \AA, respectively
--- that is, the LTE approximation tends to overestimate the derived
oxygen abundance. Such
NLTE effects are very similar to those reported by
Asplund et al. (2004), which result from full 3D+NLTE 
computations.


\subsection{The metal poor halo subgiant HD140283} \label{o_abund}

It is convenient to begin this section by giving the values of the oxygen
abundances we have obtained {\em without} assuming LTE when using
the 1D and 3D models of this metal poor star. As mentioned above,
both models are characterized by
${\rm log}g=3.7$, $T_{\rm eff}=5700$ K and [Fe/H]$=-2.5$.
The 1D model gives ${\rm log}\epsilon(\rm O)=7.11 \pm 0.06$ for
the O\,{\sc i} IR triplet, while ${\rm log}\epsilon(\rm O)=7.18$
for the [O\,{\sc i}] forbidden line at 6300 \AA. Concerning the
3D-based determination we find ${\rm log}\epsilon(\rm O)=7.08 \pm
0.06$ when using
the O\,{\sc i} IR triplet, but ${\rm log}\epsilon(\rm O)=6.96$ for
the [O\,{\sc i}] line.
Our 1D results differ slightly from those presented by
Nissen et al. (2002), probably because we have not used the same
continuum opacity package and/or identical 1D atmospheric models.
Our estimates indicate that the abundance
differences could reach $\sim 0.05-0.07$ dex.

Having given these values, it is now of interest to note that the
NLTE effects in the 1D and 3D atmospheric models of this metal
poor star are not negligible, but smaller than in the solar case
because the O\,{\sc i} IR lines originate in deeper layers where
their source functions and the level populations of their upper
levels are closer to their LTE values
(Shchukina et al. 2003).
Adopting LTE for the O\,{\sc i} lines leads to an abundance
overestimation of 0.18 and 0.16 dex for the 1D and 3D cases,
respectively. The determination of the oxygen abundance based on
the O\,{\sc i} IR triplet is nearly insensitive to 3D effects. The
correction does not exceed 0.03 dex.

On the contrary, the forbidden [O\,{\sc i}] line at 6300 \AA\ is
sensitive to the stellar atmospheric inhomogeneities. In agreement
with Nissen et al. (2002) we find that 
a 3D analysis reduces the
oxygen abundance derived from this forbidden line by $0.22$~dex.
The NLTE correction is extremely small ($<-0.01$~dex).

\section{The iron abundance} \label{fe_abund_intro}

This section focusses on the iron
abundance in the Sun and in HD140283. Our results and a
comparison with those of recent publications are given in Tables 2 and
3.


\subsection{The Sun} \label{fe_abund_sun}

Shchukina \& Trujillo Bueno (2001) applied a 3D+NLTE
approach to determine the solar iron abundance from
18  Fe\,{\sc i} lines that were carefully selected. Ideally, one would like to
determine  the iron abundance in the Sun and HD140283 employing
the same Fe\,{\sc i} line list. However, due to the relatively high solar
metallicity the solar spectral lines at wavelengths below 5000 \AA\ 
are much more affected by blends than in the metal-poor star
HD140283. As a result, overlapping wings of nearby strong lines
cause large uncertainties in
the equivalent width measurements of most of the solar Fe\,{\sc i} lines
listed in Table 1. The only exception are the lines at $\lambda >
5700$ \AA\ (i.e. only 8 Fe\,{\sc i} lines from 33 ones). 
In any case, the main reason why not the same lines can be used
in the Sun and HD140283 is that the line strengths are vastly different
in the two cases. Therefore, the weak iron lines in HD140283 are very
strong and therefore ill-suited for abundance purposes in the Sun.
Thus, to
recalculate  the iron abundance of the Sun using the lines of
Table 1 is hardly possible.

The solar  Fe\,I lines used in our 3D solar abundance
determination (see Table 1 in Shchukina \& Trujillo Bueno 2001)
were selected  from the well-known ``Kiel+Oxford''
line list according to the criteria discussed by Kostik,
Shchukina \& Rutten 1996. The lines chosen 
by Shchukina \& Trujillo Bueno (2001) account for the iron
abundance behaviour versus the lower-level excitation potential
$\chi$ -that is, for the existence of both an abundance  dependence on
$\chi$ and a sizable scatter in the abundance values for
each $\chi$. Such 18 representative Fe\,{\sc i} lines yield the
largest and lowest abundances for each $\chi$-value and spaning a large
range in $\chi$ which goes from 0 to 4.6 eV. This should be a very
suitable choice of lines in order to evaluate the largest 3D
effects. Shchukina \& Trujillo Bueno (2001) 
obtained a Fe abundance ${\rm log}
\epsilon_{\odot}$(Fe)=7.50$\pm 0.10$ for the 3D$+$NLTE case. 
The 3D+LTE approach gives a lower abundance, i.e. ${\rm log}
\epsilon_{\odot}$(Fe)=7.43$\pm 0.11$.
Neglecting the effect of granulation
inhomogeneities leads to a considerably larger value (${\rm log}
\epsilon_{\odot}$(Fe)=7.62$\pm 0.04$; e.g. Kostik et al. 1996).

We stress that the Fe\,{\sc i} lines we have used 
for the star HD140283 were
chosen to obtain a combination of lines that is comparable to 
that provided by the 18 solar
lines (concerning their equivalent widths and $\chi$ coverage). 


\subsection{The metal-poor halo subgiant HD140283} \label{fe_abund}

Like in the solar case (Shchukina and Trujillo Bueno 2001),
near-UV overionization appears to be the main NLTE mechanism for
Fe\,{\sc i} in this metal-poor star.
Both in the 1D and 3D models it causes an important
underpopulation of the Fe\,{\sc i} levels and hence a reduction in
the opacity of the Fe\,{\sc i} lines (Fig. 1, left panel). As a
result, the atmospheric heights where the line optical depth
reaches unity are shifted toward deeper layers where the line
source functions turn out to be close to the Planck function. The
relevance of this effect increases with the vertical temperature
gradient, which is
larger above the granules. For this reason, we find a strong
weakening of the Fe\,{\sc i} lines that emerge from such granular
regions, while above  the intergranular lanes the emergent
spectral line profiles do not change appreciably. On the whole,
the calculated NLTE
profiles turn out to be much weaker than when the LTE
approximation is used (see Fig. 2).
%


Concerning NLTE effects in Fe\,{\sc ii} lines it is very important
to point out that in the granular regions we find a significant
overpopulation of the excited levels of odd parity starting at
$z\rm{{^6}D^{o}}$ (Fig. 1, right panel).
According to Cram et al. (1980) and Rutten (1988) this effect is
caused by an excess in the UV  continuum radiation field near 2600
\AA, which leads to a pumping of the upper level populations of
the Fe\,{\sc ii} lines. As a
consequence, the source functions of the Fe\,{\sc ii} lines with
high excitation energy of the upper level
(${\chi}_{\rm up} \ge 4.8$~ev) exceed significantly the Planck
function values. This implies that such Fe\,{\sc ii} lines are
weaker than stipulated by the LTE approximation and may even be in
emission instead of in absorption. This radiative pumping
mechanism takes place in both the 1D and 3D models. Like in the
case of
Fe\,{\sc i} lines it is sensitive to the vertical temperature
gradient. The steeper the gradient, the larger the pumping. As a
result, in the granular regions where the temperature gradients
are larger the effect is much more pronounced (see Fig. 1).


The importance of such an effect is illustrated in Fig.~3, which
shows both NLTE and LTE results for two Fe\,{\sc ii} lines using
the 3D model atmosphere of the metal-poor subgiant HD140283. We
point out that emission is much more pronounced for the emergent
profiles that originate above the stellar granular regions, while
above  the intergranular lanes the spectral line profiles are
hardly modified with respect to the LTE case. Allowing for
departures from LTE
produces a drastic change in the
calculated spatially
averaged flux of
the weak Fe\,{\sc ii} $\lambda$6247.56 \AA\ line
in the 3D model: the absorption
profile is replaced by one in emission.
Note, that this spectral line is taken from the set of weak
Fe\,{\sc ii} lines that Nissen et al. (2002) assumed to be
insensitive to NLTE effects.
However, we find that such type of lines do suffer from the
above-mentioned UV pumping mechanism because all of them belong to
the aforementioned category of lines, i.e. their upper levels are
of odd parity and are situated above the
$z\rm{{^6}D^{o}}$-level.
The stronger Fe\,{\sc ii} $\lambda$5316.62 \AA\ line displayed in
Fig.~3 is also from this category. Figure 3 clearly demonstrates
that the stronger the spectral line the smaller the relative
effect of the UV-pumping. For this line emission arises only at
the core of the ``granular'' profiles. Spatial averaging masks
completely such an emission, but makes the flux profile weaker
than in LTE. It is also of interest to mention that in the 1D
stellar atmosphere model departures from LTE produce similar
changes in the Fe\,{\sc i} and Fe\,{\sc ii} lines but they are
less pronounced than in the 3D case. For instance, 
the flux profile of the weak Fe\,{\sc ii} $\lambda$6247.56 \AA\ line
turns out to be in absorption for the 1D+NLTE case.


How large could be the impact of UV-overionization and
UV-overexcitation on the iron abundance determination in the star
considered?
It is obvious that in order to compensate for a weakening of the
emergent spectral line profiles in the stellar atmosphere
model under consideration the iron abundance must be
increased in comparison with the LTE case. Figure~4 quantifies
this conclusion.
It  shows the LTE versus NLTE iron abundances derived from either
Fe\,{\sc i} or Fe\,{\sc ii} lines in the 3D and 1D stellar
atmospheric models of the metal poor star HD140283.
As seen in the figure, the NLTE abundance corrections for the
Fe\,{\sc i} lines are extremely large in the 3D model ($\sim0.9$
dex) and a bit smaller in the 1D case ($\sim 0.6$ dex). On
the other hand, it is very important to point out that NLTE
effects on the iron abundance are most pronounced for the weak
Fe\,{\sc ii} lines. The four lines
used by Nissen et al (2002) for their iron abundance
determinations
reveal strikingly large deviations from LTE.
On average, the NLTE abundance corrections for the Fe\,{\sc ii}
lines increase from $\sim 0.16$ dex in the 1D case to $\sim
0.4$~dex in the 3D model.


When NLTE effects are taken into account we find no difference
between the mean Fe\,{\sc i}-based abundances obtained via the 1D
and 3D models (${\rm log}\epsilon(\rm Fe){\approx}5.77$).
Therefore, when NLTE effects are taken into account the Fe\,{\sc
i} lines formed in the 3D model are hardly sensitive to
granulation inhomogeneities.
%
However, for the Fe\,{\sc ii} lines the 3D abundance correction
(3D+NLTE $-$ 1D+NLTE) is significant --- that is, 0.26~dex.

When the LTE approximation is used the mean iron abundance
obtained via spectrum synthesis of Fe\,{\sc i} lines in the 3D
model is lower than for the 1D case (cf. Asplund et al. 1999;
Nissen et al. 2002). The differences between the
1D and 3D determinations depend on the lower excitation potential
$\chi$ of the spectral line under consideration. They range from
$\sim 0.6$ dex at $\chi \simeq 0$~eV  to  $\sim 0.2$ dex at $\chi
\sim 2$~eV, while the mean difference is smaller ($\sim 0.3$ dex).
In contrast, on average the Fe\,{\sc ii} lines are insensitive to
3D effects when LTE is assumed.



\section{Revising the stellar parameters} \label{discussion}


There are several discrepancies indicating that the iron abundance
obtained in Sect.~\ref{fe_abund} for the star HD 140283 cannot be
considered as reliable, neither in the 1D case nor in the 3D case.
For example, we have (1) a divergence of the average abundances
derived from both ionization stages, (2) a correlation of the
abundances derived from Fe\,{\sc i} lines with the lower-level
excitation potential and (3) a large scatter in the abundances
derived from Fe\,{\sc ii} lines (particularly, the weak ones).
The elimination of any of such discrepancies cannot be simply
achieved in terms of uncertainties in observed equivalent widths,
oscillator strengths, inelastic electron
collisional rates and photoionization cross sections. 
Our numerical experiments show that the typical errors
introduced by uncertainties in such
quantities do not exceed 0.1 dex (see Shchukina and Trujillo Bueno
2001). The iron abundance error that results from neglecting
the UV haze opacity for the metal-poor star HD140283
is even smaller than 0.1 dex. In our opinion, a reasonable possibility for
explaining the reported discrepancies is to
revise  the stellar parameters of the subgiant HD140283.

In this section we aim at finding the ``best-choice'' of stellar
parameters for the star HD140283, both for the NLTE and LTE cases.
It is important to emphasize from the outset that in our NLTE
computations we have neglected inelastic collisions with neutral hydrogen
atoms for excitation and ionization. At present, there are no
reliable values for the ensuing collisional rates. 
The often-used classical Drawin (1968) formula leads 
to very uncertain estimates and to a substantial overestimation 
of the collisional rates when detailed quantum mechanical 
and/or laboratory data is available, such as for Li and Na 
(see Biberman et al. 1987; Lambert 1993; Holweger 1996; Barklem
et al. 1998, 2003; Belyayev et al. 1999; Shchukina \& Trujillo
Bueno 2001; Asplund et al. 2004),
which explains why the conclusions of Gratton et al. (1999)
have been critized. Obviously,
inelastic collisions with neutral hydrogen atoms would tend to
reduce the NLTE effects. 
Therefore, the difference between the NLTE and LTE 
iron abundances may be considered as the maximum effect 
that inelastic collisions might produce. In our opinion, the tentative
conclusion of Korn, Shi \& Gehren (2003), achieved via a 1D+NLTE
analysis, that hydrogen collisions might be efficient in the atmosphere of HD140283 needs to be re-investigated in the light of 3D hydrodynamical
models because, as pointed out
by these authors, the lower temperatures
of the 3D models would alleviate their need of invoking thermalizing collisions
to counterbalance photoionization.

Our analysis for the revision of the stellar parameter is based on
the following criteria:

i) the average abundances obtained from Fe\,{\sc i} and
Fe\,{\sc ii} lines have to be equal.

ii)  the ``best choice'' solution has to give the minimum mean
standard deviation ($\sigma$) for the abundances derived from
Fe\,{\sc i} and Fe\,{\sc ii} lines.

iii)  
abundances have to be independent of the lower excitation
potential ($\chi$). We restrict ourselves to revise only the
metallicity and the effective temperature. The given surface
gravity of the star HD140283 is reliable, since it has been
obtained from accurate {\it Hipparcos} parallaxes.

Figure~5  helps us to find the ``best-choice'' of stellar
parameters for the star HD140283. The left panels show the
variation of the Fe\,{\sc i} and Fe\,{\sc ii} abundances with
metalicity, $T_{\rm eff}$, and ${\rm log}g$ in a suitable grid of
1D model atmospheres.
Our results clearly demonstrate that changes in $T_{\rm eff}$
seriously affect the Fe\,{\sc i}-based abundances. However, such
changes in $T_{\rm eff}$ are unimportant for abundance
determinations from Fe\,{\sc ii} lines. On the other hand, the
Fe\,{\sc ii}-based abundances are slightly sensitive to the
changes in ${\rm log}g$, which is not the case for the abundances
obtained from Fe\,{\sc ii} lines. In addition, the dependence on
metallicity is more pronounced for Fe\,{\sc i} lines, while for
Fe\,{\sc ii} lines there is no such a variation. It is interesting
to note that when LTE is assumed, then only the correlation of the
inferred Fe\,{\sc i} abundances with $T_{\rm eff}$ is present.

\subsection{Stellar parameter determinations: 1D case} \label{param_1D}


Consider first the 1D+LTE modeling case for the iron lines.
Figure~5 (middle panel to the {\em lhs}) and Fig. 4 (bottom panel
to the {\em rhs}) show that in this case identical abundances from
both Fe\,{\sc i} and Fe\,{\sc ii} lines could be obtained around
$T_{\rm eff}$ close to 5700 K. In fact, our detailed analysis
gives $T_{\rm eff} \simeq 5670$ K and [Fe/H]$\simeq -2.3$. With
these stellar parameters Fe\,{\sc ii}-based abundances turn out to
be insignificantly scattered ($\sigma \sim 0.1$ dex) around the
average value. However, the abundances derived from Fe\,{\sc i}
lines show a sensitivity to the lower-level excitation potential
($\chi$) that is similar to the trend seen in the bottom panel to
the {\em rhs} of Fig. 4.


Secondly, consider the 1D+NLTE modeling case. In order to reach a
good agreement between NLTE abundances derived from
Fe\,{\sc i} and Fe\,{\sc ii} lines we would have to reduce the
effective temperature down to $T_{\rm eff}=5450$ K and to increase
the stellar metallicity up to  [Fe/H]$\simeq -2.1$. With such
stellar parameters the $\chi$-dependence of the
Fe\,{\sc i} abundances vanishes. However, the abundances derived
from Fe\,{\sc ii} lines become undesirably spread around the mean
value ($\sigma \sim 0.3$ dex).


\subsection{Stellar parameter determinations: 3D case} \label{param_3D}

In order to achieve a reasonable estimation of the ``best choice''
of stellar parameters the curves of Fig.~5 (left panels) have to
be corrected for the 3D effects of granulation inhomogeneities.
This means that we need to know how the 3D abundance corrections
change with [Fe/H], $T_{\rm eff}$ and ${\rm log}g$ for both the
LTE and NLTE cases. Unfortunately, a suitable grid of 3D models
for several effective temperatures and metalicities similar to
that provided by Kurucz for the 1D case is not yet available,
besides the fact that NLTE computations in 3D are presently very
computationally costly. However, we can try to estimate the true stellar
parameters of the star HD140283 via the following strategy, which
is inspired by our conclusion that 3D effects are important for
the Fe {\sc i} lines only if LTE is assumed and that 3D effects
are relevant for the Fe {\sc ii} lines only when we allow for
departures from LTE.

Firstly, let us choose the 3D+LTE modeling approach. As we have
seen, if LTE is assumed then the Fe {\sc i} lines are sensitive to
the 3D effects while the Fe {\sc ii} lines are not. Therefore, if
the LTE approximation is used it is much safer to revise the
stellar parameters via iron abundances derived from Fe {\sc ii}
lines. We find $T_{\rm eff} \approx 5700$ and [Fe/H]$\approx -2.3$
(see the bottom panel to the {\em rhs} of Fig.~4).

Secondly, consider the 3D+NLTE approach. It is obvious that in
this case we better avoid the use of Fe {\sc ii} lines because we
have shown that they are sensitive to both 3D and NLTE effects.
However, we have demonstrated also that at least for the available
3D model of the star HD140283 the NLTE iron abundance derived from
Fe {\sc i} lines is very similar to that obtained via the 1D+NLTE
modeling approach. For this reason, we think that we can estimate
the true stellar parameters by choosing Fe {\sc i} lines and a
1D+NLTE modeling approach. We find that the standard deviation
$\sigma$ minimizes at $T_{\rm eff} \simeq 5600$ K and
[Fe/H]$\simeq -2$. Now the mean iron abundance turns out to be
${\rm log}\epsilon(\rm Fe)=5.54 \pm 0.14$. Moreover, the
dependence of the derived iron abundances with $\chi$ becomes
negligible.


\subsection{Uncertainties in the determination of the stellar parameters
and the oxygen abundance} \label{param_ox}

As we have just seen, depending on the approach used for modeling
the iron lines we get four ``best choice'' sets for the stellar
parameters of the star HD140283, with $T_{\rm eff}$ in the range
$\simeq 5450-5700$ K and [Fe/H]  between $-2.0$ and $- 2.3$ (see
Table 4).
How large could be the impact of such uncertainties in the stellar
parameters of the star HD140283 on the determination of the oxygen
abundance?
The right panels of Fig.~5 clarify this issue. We can see that
when 1D models are used then both the O\,{\sc i} and
[O\,{\sc i}]-based abundances are sensitive to  changes in
$T_{\rm eff}$, but not too much to changes in ${\rm log}g$ and/or
in metallicity. The sensitivity of the NLTE abundance corrections
to these parameters (${\rm log}g$ and metallicity) is much
smaller. On average, they vary between 0.17 and 0.19 dex  for
O\,{\sc i} lines, while it is negligible for the [O\,{\sc i}] line
($<0.01$ dex). Our 1D+NLTE modeling of oxygen lines shows that
abundances derived from O\,{\sc i} and [O\,{\sc i}] lines give the
same value (i.e., ${\rm log}\epsilon(\rm O)=7.15$) for $T_{\rm
eff}=5650$ K. We now point out that among our  ``best choice''
sets of stellar parameters only one has a $T_{\rm eff}$-value
($T_{\rm eff}=5450$ K) which considerably deviates from that with
5650 K. Such relatively low value for the effective temperature
(i.e., $T_{\rm eff}=5450$ K) causes a large disagreement between
the NLTE O\,{\sc i} and [O\,{\sc i}]-based abundances amounting to
$\sim 0.34$ dex. Moreover, as we have indicated above the set of
parameters $T_{\rm eff}=5450$ K and [Fe/H]$=-2.1$ produces an
unpleasant scatter of the Fe\,{\sc ii}-based abundances around the
mean value. In view of these factors we think reasonable to
eliminate such particular set of stellar parameters.

The remaining three sets  of stellar parameters,
i.e. $T_{\rm eff}=5670$ K \& [Fe/H]$=-2.3$, $T_{\rm eff}=5700$ K
\& [Fe/H]$=-2.3$, and $T_{\rm eff}=5600$ K \& [Fe/H]$=-2.0$,
correspond to the case of 1D+LTE modeling for both  Fe\,{\sc i}
and Fe\,{\sc ii} lines, to 3D+LTE modeling of Fe\,{\sc ii} lines,
and to 3D+NLTE modeling of Fe\,{\sc i} lines, respectively.
It is important to note that this range of effective temperatures
is similar to the typical uncertainty in the determination of
$T_{\rm eff}$ via the application of the infrared flux method.
The small differences in the effective temperature
 implies a rather small oxygen abundance variation.
In particular, our 1D+NLTE oxygen abundance determination from the
O\,{\sc i} IR lines using the former set gives ${\rm
log}\epsilon(\rm O)=7.13$, while the latter set leads to ${\rm
log}\epsilon(\rm O)=7.20$. For the [O\,{\sc i}] line we get ${\rm
log}\epsilon(\rm O)=7.16$ and ${\rm log}\epsilon(\rm O)=7.11$,
respectively.


We must consider now how the dependencies shown for the oxygen
lines in the {\em rhs} panels of Fig.~5 are going to be modified
due to the effect of granulation inhomogeneities. As mentioned
above, a major obstacle to investigate this point is the lack of a
suitable grid of 3D models. In any case, this 3D issue concerns
especially the [O\,{\sc i}] line given its significant sensitivity
to granulation inhomogeneities. According to Nissen et al. (2002)
the main effect is due to differences in the continuum opacities
between the 1D and 3D models. Since the opacity is provided
primarily by
$\rm H^{-}$ atoms
the strength of the weak [O\,{\sc i}] line turns out to be
inversely proportional to electron density. In the 3D
hydrodynamical model of the star HD140283, which has been obtained
assuming $T_{\rm eff}=5700$ K and [Fe/H]$=-2.5$, the deficit of
electrons is particularly large in the cool surface layers of the
granular regions, while it is less pronounced in both the
intergranular plasma and in the 1D model. As a result, the
spatially averaged equivalent width of the [O\,{\sc i}] line in
such a 3D model increases and, consequently, the derived oxygen
abundance decreases in comparison with the 1D modeling result.
Obviously, the 3D correction obtained in Sect.~\ref{o_abund} for
this forbidden line assuming that the stellar parameters are
$T_{\rm eff}=5700$ K and [Fe/H]$=-2.5$ cannot be used for the
3D+NLTE case with lower effective temperature ($T_{\rm eff}=5600$
K) and higher metalicity ([Fe/H]$=-2.0$).

As pointed out by Asplund et al. (1999)
the deficit in the number density of electrons depends mainly on
the metalicity. 3D hydrodynamical simulations for higher [Fe/H]
values will have larger surface temperatures and electron
densities. Thus, one can expect that at [Fe/H]$=-2$ the 3D
abundance correction for the [O\,{\sc i}] line has to be smaller.
Obviously, the {\em exact} value of such corrections will have to
await the development of a new 3D model for the
revised stellar parameters.
It should be noted that such a sensitivity to 3D effects may be
assumed also for the UV lines of OH. However, the impact on
3D abundance corrections when changing the metallicity from $-2.5$
to $-2.0$ seems to be rather small if we suppose that the results
obtained by Asplund \& Garc\'\i a P\'erez (2001) for dwarfs can
be applied to the subgiant HD140283 (see their Table
2, which shows that the 3D effects on the UV lines
of OH is $\sim -0.5$ at
[Fe/H]$=-2$ and  $\sim -0.6$ at [Fe/H]$=-3$). The additional
conclusion that follows from their Table 2 is a high sensitivity of
the oxygen abundances derived from the OH molecules to the
metallicity of the star. As a result, a 3D model with
higher metallicity ([Fe/H]$=-2$) might bring the oxygen abundances
derived from molecules
(${\rm log}\epsilon(\rm O)=7.20$ for $T_{\rm eff}\approx 5800$) to
closer agreement with the abundance  derived from the O\,{\sc i}
triplet   ( ${\rm log}\epsilon(\rm O) \approx 7.20$ for $T_{\rm
eff}=5600$ K.)

%

An extra problem is related with observations of the [O\,{\sc i}]
line in the star HD140283. According to
Nissen et al. (2002) this weak line ($W
\simeq 0.5$ m\AA) is blended with a stronger
$\rm H_{2}O$ line ($W \approx 1$ m\AA). 
These authors estimated that
the available equivalent width measurements have an uncertainty of
$\pm 0.2 - \pm 0.3$~m\AA\ . Such an uncertainty is very large,
because it is about 50\% of the observed equivalent width, and implies an uncertainty of $\pm 0.13 - \pm
0.19$~dex in the derived oxygen abundance (Fig.~5, middle right
panel).

Alternatively, in a first approximation we could neglect the
sensitivity of the O\,{\sc i} lines to the granulation
inhomogeneities. Furthermore, the observation errors for these
lines in HD140283 are relatively small.
For example, the observed equivalent widths of the O\,{\sc i}
triplet that have been published in the literature
(Abia \& Rebolo 1989;
Tomkin et al. 1992;
Nissen et al. 2002;
G. Israelian, private communication),
and which we have used in our investigation, differ by about $\pm
1$~m\AA. Their impact on the derived oxygen abundance lies in the range
$\pm 0.06$~dex (Fig.~5, middle right panel).
It is clear that the O\,{\sc i} lines seem to be the most reliable
lines for deriving the oxygen abundance in the star HD140283.

Therefore, when the stellar parameters are revised neglecting 3D
and NLTE effects for the iron lines, and when the oxygen abundance
derived from the O\,{\sc i} lines  is corrected for NLTE effects,
then the mean value of the oxygen-to-iron ratio seems to be
[O/Fe]$\simeq 0.7$  at  [Fe/H]=$-2.3$. This value is in good
agreement with the quasi-linear trend of [O/Fe] vs. [Fe/H]
obtained by
Nissen et al. (2002)
using standard 1D modeling. However, the key point to keep in mind
is that a 3D+NLTE modeling of the O\,{\sc i} IR triplet
(using the 3D+NLTE stellar parameters derived from the Fe\,{\sc i}
lines)
would give [O/Fe]$\simeq 0.5$ at [Fe/H]=$-2.0$.
Obviously,
the analysis of a single metal-poor star is not enough
to opt for one of the following two possibilities:
(a) a linear rise in the [O/Fe] ratio
vs. metallicity (see, e.g., Fig. 11 of Israelian et al. 2001)
or (b) a
plateau between 0.4 and 0.5 dex in the metallicity
range -3${<}$[Fe/H]${<}-1$.
\footnote{We point out that with the stellar parameters used by
Asplund et al. (1999) for the 3D modelling of the star HD140283
(i.e., $T_{\rm eff}=5700$ and [Fe/H]=$-2.5$) the NLTE oxygen
abundance
(${\rm log}\epsilon({\rm O})=7.08\pm 0.06$)
derived  from the  O\,{\sc i} IR triplet and
the NLTE Fe\,{\sc i}-based iron abundance
(${\rm log}\epsilon({\rm Fe})=5.77\pm 0.186$) leads to a small
oxygen-to-iron ratio ([O/Fe]$\sim 0.1$ dex).
Moreover, there is no overabundance of oxygen at all
if one uses the [O\,{\sc i}]-based abundance.
It is also of interest to mention that in our preliminary study
(see Shchukina et al. 2003) we obtained a slightly larger value
for [O/Fe] because we used a smaller number of Fe\,{\sc i} lines.}
However, we think that our investigation indicates the ``road to be taken''
towards the resolution of the problem.
 
%



\section{Conclusions} \label{conclusion}

We have derived the solar oxygen abundance from the O\,{\sc i} IR
triplet using Asplund's et al. (2000) 3D hydrodynamical model of
the solar photosphere and a NLTE modeling approach. We find ${\rm
log} \epsilon({\rm O})=8.70\pm 0.06$, which is in good agreement
with the determinations of Allende Prieto et al. (2001), Shchukina
et al. (2003) and Asplund et al. (2004). Actually, if
for the O\,{\sc i} IR triplet we use
the same oscillator strengths adopted by Asplund et al. (2004) we
then obtain a value
very similar to that
reported by them
(i.e., ${\rm log}\epsilon({\rm O})=8.65$).



In agreement with previous investigations, we confirm that LTE
synthesis of Fe\,{\sc i} lines in the 3D hydrodynamical model of
this star gives flux profiles significantly different from those
obtained in a 1D model for the same stellar parameters. However,
when the LTE approximation is used for the synthesis of Fe\,{\sc
ii} lines we find negligible differences between the 1D and 3D
cases.

The differences between the iron abundances obtained from
Fe\,{\sc i} lines when assuming LTE in the 1D and 3D models of the
star HD140283 turn out to depend on the lower excitation potential
of the spectral line under consideration. The NLTE abundance
corrections for Fe\,{\sc i} lines are significant, especially for
the low-excitation lines in the 3D model.

We have three particularly important results concerning the
determination of the iron abundance in metal poor stars like
HD140283:

(1) If NLTE effects are
taken into account when synthetizing the Fe\,{\sc i} lines in the
1D and 3D models of this star, then the derived iron abundance
turns out to be practically the same in both cases.

(2) Contrary to a generally accepted belief, the Fe\,{\sc ii}
lines turn out to be significantly affected by NLTE
effects. In particular, a full NLTE modeling should be carried out
for the Fe\,{\sc ii} lines with upper levels of odd parity
starting from $z\rm{{^6}D^{o}}$, and especially the weaker lines.

(3) The iron abundance of the star HD140283 is estimated to
be $\sim 0.5$ dex higher than previously thought.

For the O\,{\sc i} IR triplet the NLTE correction of the oxygen
abundance in the 1D and 3D models is practically the same and less
than 0.2 dex. Such a NLTE correction is negligible for the
[O\,{\sc i}] line.
%
We have confirmed that the mean value of the oxygen abundance
derived from the O\,{\sc i} IR lines is insensitive to 3D effects,
while such effects reduce the abundance derived from the
[O\,{\sc i}] line by $\sim 0.2$ dex.


With the here presented NLTE calculations for
Fe and O, both the 1D and 3D models
lead to inconsistencies in the iron and oxygen
abundance determinations.
The discrepancies in the derived iron and oxygen abundances cannot
be removed by taking into account NLTE and/or 3D effects. This
result has led us to investigate whether these
discrepancies can be resolved by a modification of the stellar
parameters of the star HD140283.
We have shown that previous studies may have underestimated the
metallicity of this star and overestimated its effective
temperature. We find [Fe/H]$\simeq-2.0$ (instead of
[Fe/H]$\simeq-2.5$) and $T_{\rm eff}=5600$
(instead of $T_{\rm eff}=5700$).
%
With these new stellar parameters the iron and oxygen abundances
in the star HD140283 would be
${\rm log}\epsilon({\rm Fe})=5.54\pm 0.14$ and
${\rm log}\epsilon({\rm O})=7.20\pm 0.06$,
respectively. Taking into account  the  {\it low} value for the
solar oxygen abundance
(i.e., ${\rm log}\epsilon_{\odot}$({\rm O})=8.70$\pm 0.06$) and
our previously determined value of the solar iron abundance
(i.e., ${\rm log}\epsilon_{\odot}$({\rm Fe})=$7.50\pm 0.10$; see
Shchukina \& Trujillo Bueno 2001) we
find that the oxygen-to-iron abundance ratio is
${\rm [O/Fe]}\approx 0.5$ at  [Fe/H]=$-2$.

Besides the reported NLTE effects for the Fe {\sc ii} lines,
our main conclusion here is that the metalicity of this type of
metal-poor stars might well be significantly larger than
previously thought. Obviously,
the analysis of a single metal-poor star is not enough
to fully resolve the puzzling behavior of the O/Fe ratio
in metal-poor stars.
However, we think that the present investigation
at least indicates the road to be taken.


\acknowledgments

We are very grateful to Ram\'on Garc\'\i a L\'opez and
Garik Israelian
for allowing us to use their stellar observations of iron and
oxygen lines.
We are also grateful to
I. Vasiljeva for helping us with the radiative transfer
computations during the early stages of this project.
This work has been funded by the European Commission through INTAS
grant 00-00084 and by the Spanish Ministerio de Educaci\'on y Ciencia through project AYA2001-1649.





 \clearpage



\begin{figure*}
 \plotone{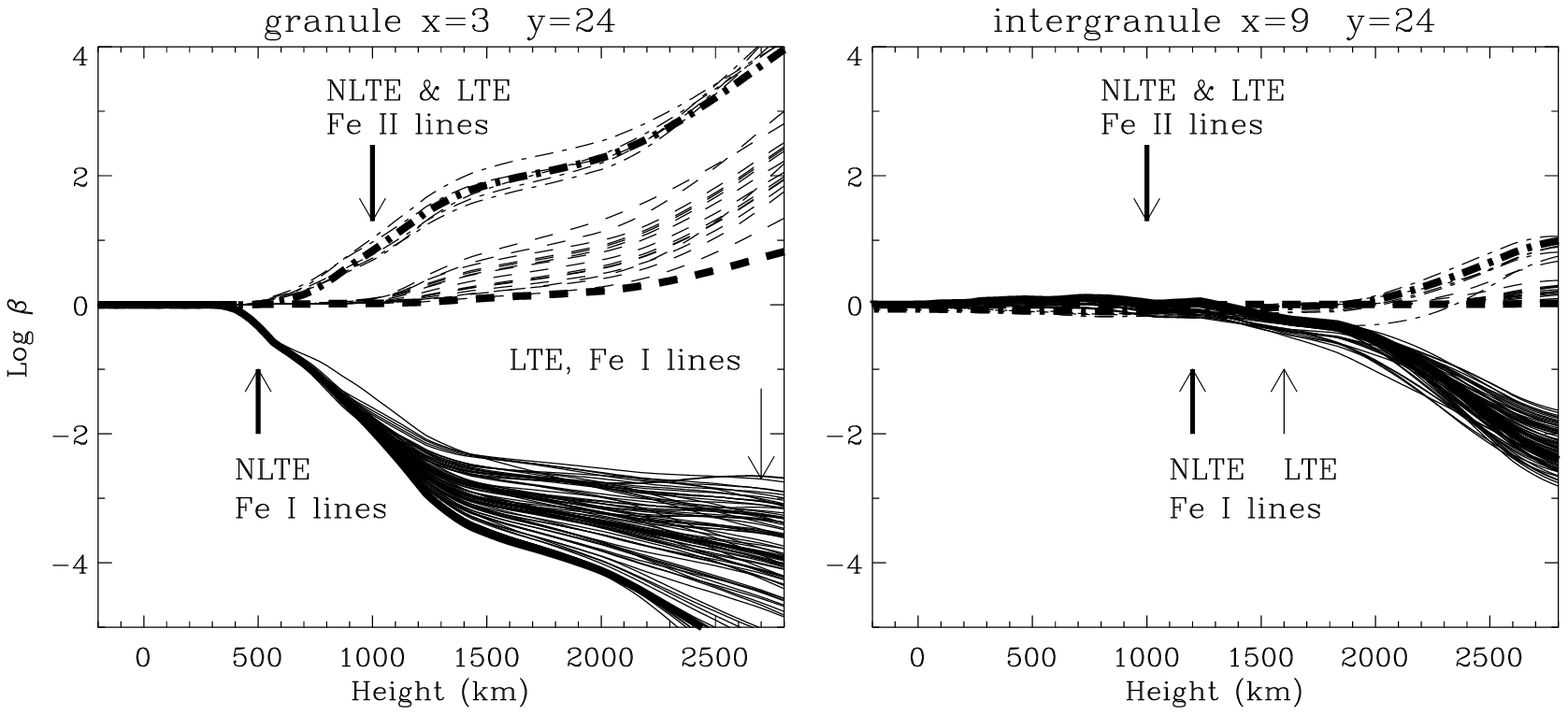}
\caption{Departure coefficients $(\beta)$ vs. height in the 3D
atmospheric model of the metal-poor star HD140283.
{\it Left panel:} Representative ``granular'' point.
{\it Right panel:} Representative ``intergranular'' point. The
solid lines indicate the $\beta$-coefficients for the Fe\,{\sc i}
levels with excitation potentials below 5 eV. The dashed and
dashed-dotted lines refer to the $\beta$-coefficients of the lower
and upper  levels of the Fe\,{\sc ii} lines in Table 1,
respectively. The thick lines correspond to
the Fe\,{\sc i}  ground level $a\rm{{^5}D}$,
to the Fe\,{\sc ii}  ground level $a\rm{{^6}D}$, and to
the Fe\,{\sc ii}  excited level of odd parity  $z\rm{{^6}D^{o}}$,
respectively. Arrows mark the representative atmospheric heights
below which one of the Fe\,{\sc i} lines ($\lambda$5429.699\,\AA)
and the Fe\,{\sc ii} lines from Table 1 originate in the NLTE and
LTE cases. Note that such NLTE and LTE heights are nearly the same
for the Fe\,{\sc ii} lines.
\label{figb}}
\end{figure*}

\begin{figure*}
\plotone{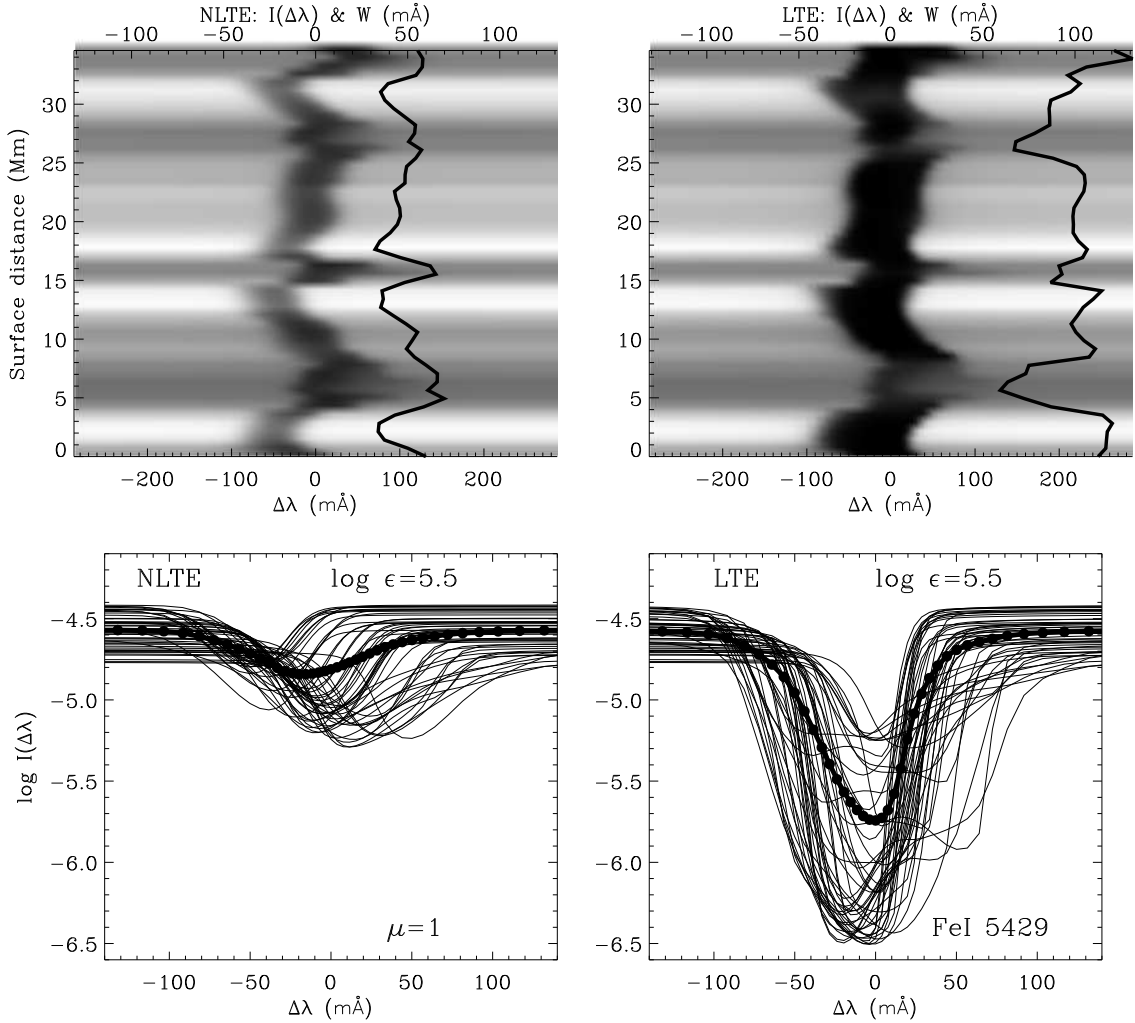} \caption{The emergent disk-center
intensity profiles of the Fe\,{\sc i} line $\lambda$5429.699\,\AA\
calculated in the 3D hydrodynamical model of the metal-poor star
HD140283. {\it Left panels} correspond to the NLTE case, while the
{\it righ panels} to LTE. {\it Top  panels:} the wavelength
variation of the emergent intensity that one would be able to
extract at each point along the spectrograph's slit if the
`surface' of the star HD140283 could be spatially resolved. The
``wiggles'' are caused by the Doppler shifts associated with the
velocity flows of the stellar granulation pattern.      Bright
strips correspond to granules, while  dark ones to intergranules.
The hypothetical slit is
located along the position Y=24 in the 3D model. Intensities are
given in absolute energy units
(erg cm$^{-2}$s$^{-1}$ster$^{-1}$Hz$^{-1}$). Black lines show
variations of the line  equivalent widths $W$ along the slit. {\it
Bottom panels:} The spatially resolved emergent intensity profiles
at each of the 50 surface gridpoints considered. The thick solid
lines with filled circles show the resulting spatially averaged
intensity. \label{fig0}}
\end{figure*}

\clearpage
%

\begin{figure*}
\plotone{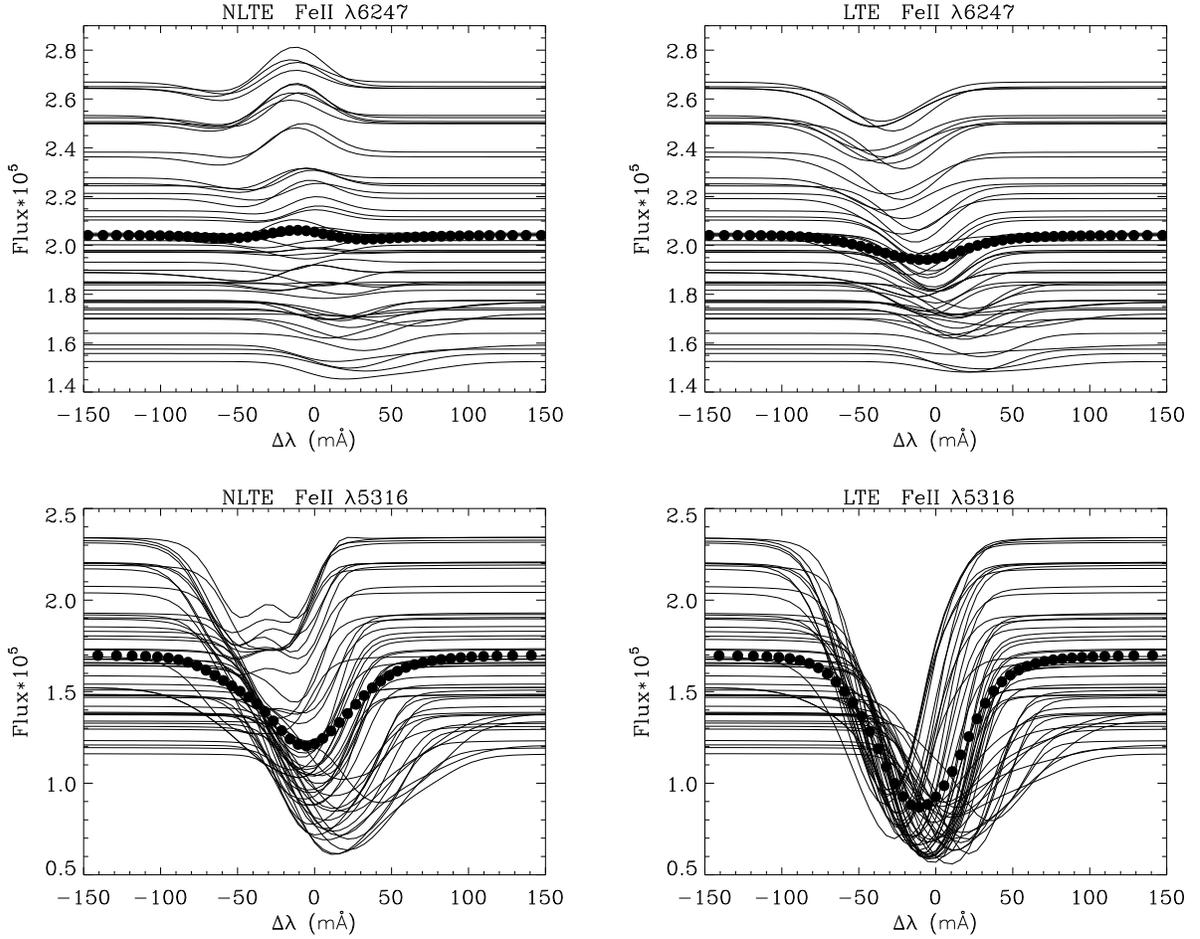}
\caption{
The flux profiles of the weak
%
Fe\,{\sc ii} line at 6247.56 \AA\ ({\it top panels})
and of  the moderately strong
%
%
Fe\,{\sc ii} line at 5316.62 \AA\ ({\it bottom panels}) calculated
in the 3D hydrodynamical model of the metal-poor star HD140283
(${\rm log}g=3.7$, $T_{\rm eff}=5700$ K and [Fe/H]$=-2.5$).
 The profiles  were synthesized  using
 the  iron abundance corresponding to the metallicity log$\epsilon$=5.0.
{\it Left panels} show the results for the  NLTE case while the
{\it right panels } refer to the LTE calculations. Individual thin
curves show the computed emergent fluxes for the ``granular'' and
``intergranular'' models corresponding to the 50 spatial grid
points situated along one of the $Y$-directions in the 3D model
atmosphere of the metal-poor subgiant HD140283. The thick solid
line with filled circles in each of the panels shows the resulting
spatially averaged flux. Fluxes are given in absolute energy units
(erg cm$^{-2}$s$^{-1}$Hz$^{-1}$). The profiles associated with the
dark intergranular lanes of the stellar granulation have a
redshift and a lower continuum flux. Bright granules are
characterized by a higher continuum flux and a blue line shift.
Note that the flux profile of the Fe\,{\sc ii} line at
6247.56 \AA\ is observed in absorption ($W>0$) while the
synthesized NLTE  flux profile turns out to be in emission
(calculated $W<0$).
\label{fig1}}
\end{figure*}
%

\clearpage

\begin{figure*}
\plotone{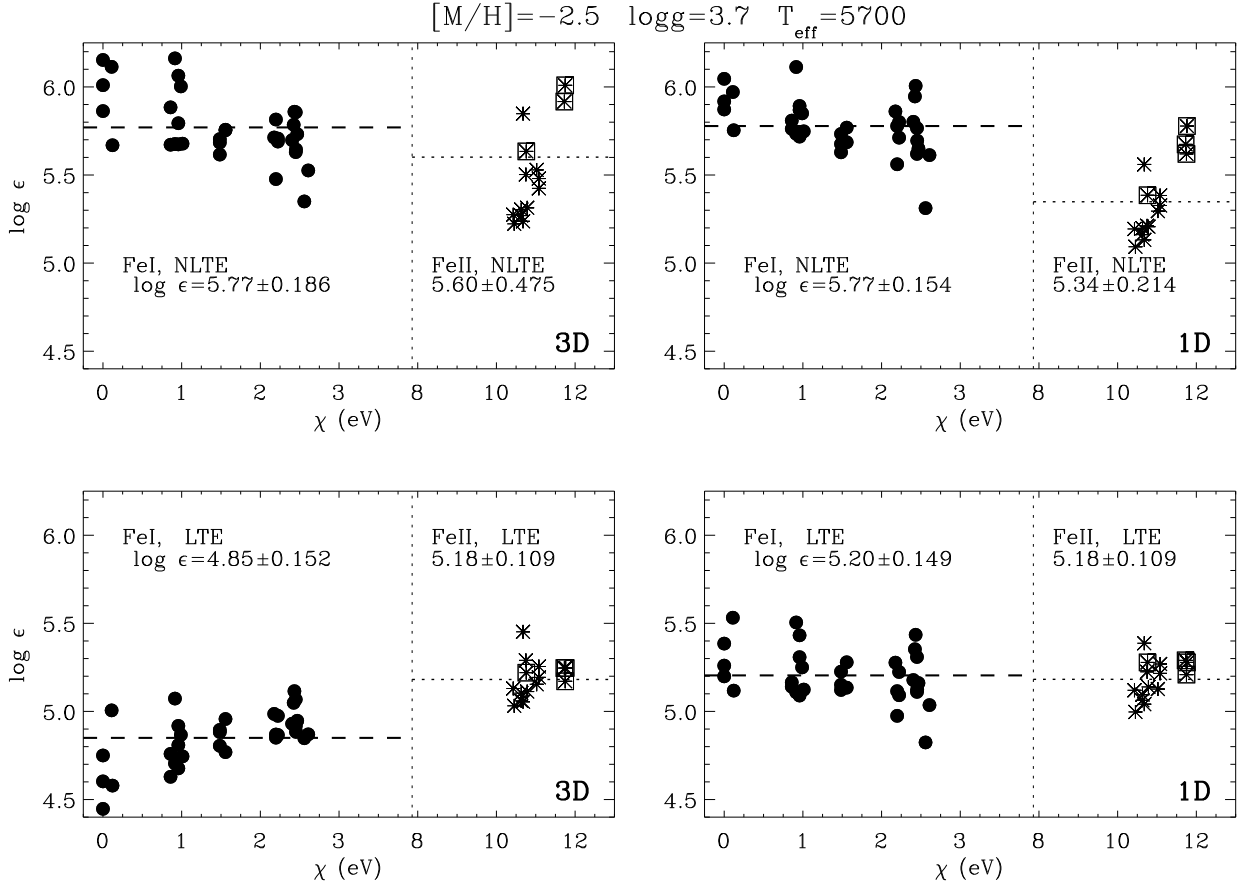}
\caption{ The iron abundances derived  from  Fe\,{\sc i} lines
(filled circles) and from Fe\,{\sc ii} lines (stars) vs. the lower
level excitation potential of the chosen spectral lines. {\it Left
panels} show the results for the 3D model atmosphere of the star
HD140283 while the {\it right panels} for the 1D model. {\it Top
panels}: NLTE. {\it Bottom panels}: LTE. The four weak Fe\,\,{\sc
ii} lines (${\lambda}{\lambda} 6149.238, 6247.562, 6432.654, 6456.
376$ m\AA) used by Nissen et al. (2002) are  marked with squares.
The dashed and dotted horizontal lines indicate the  mean
abundances found from the set of Fe\,{\sc i} and Fe\,{\sc ii}
lines, respectively. The corresponding values of the mean
abundances and their standard deviation are given inside the
panels. \label{fig2}}
\end{figure*}

\clearpage

\begin{figure*}
\plotone{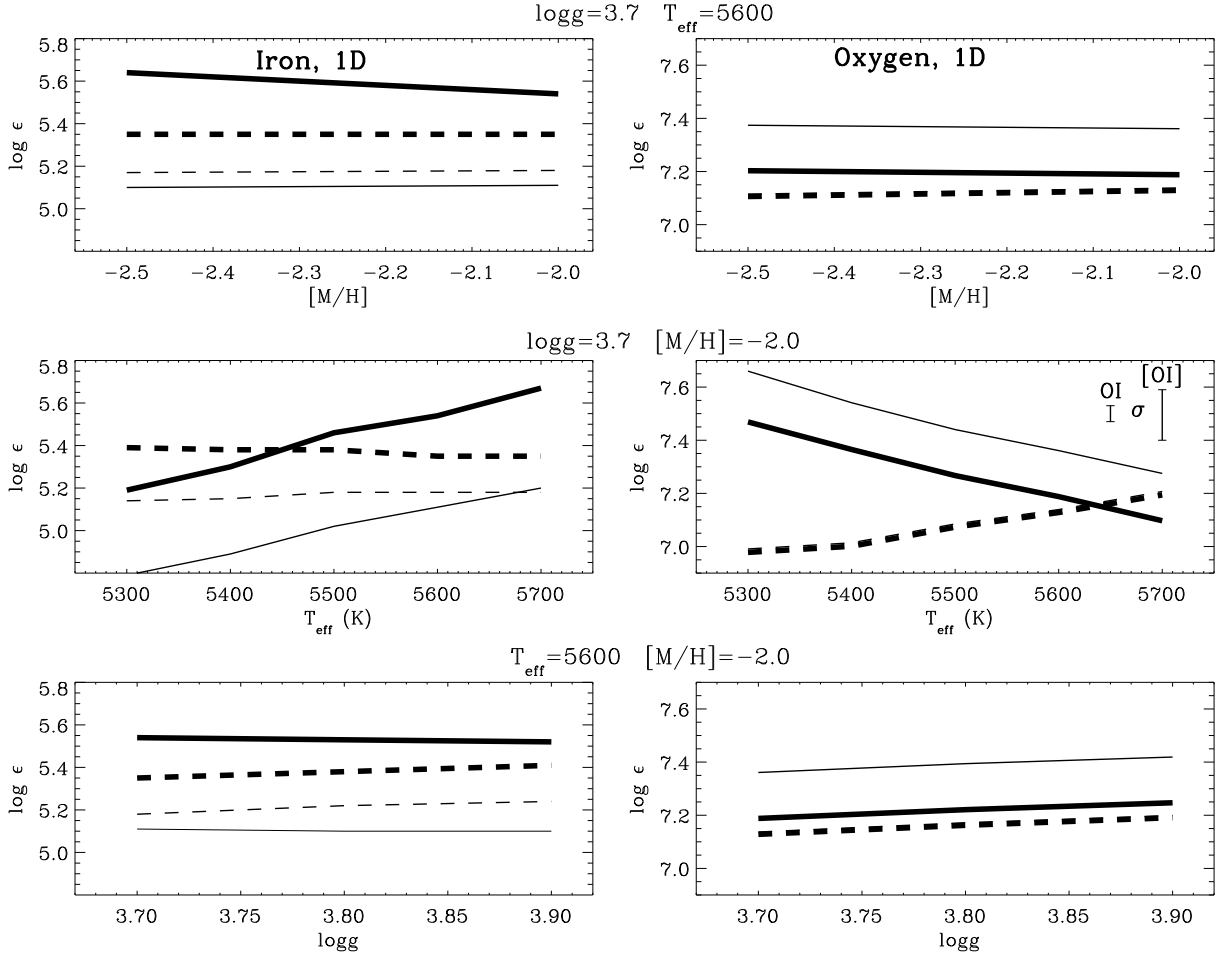} \caption{The computed abundances vs.
the metallicity ({\it top}), vs. the effective temperature ({\it
middle}), and vs. the stellar surface gravity ({\it bottom}) for
the set of 1D stellar model atmospheres. {\it Left panels}:
results for iron, with the solid and dashed  lines indicating the
mean Fe\,{\sc i}- and Fe\,{\sc ii}-based abundances, respectively.
{\it Right panels}: results for oxygen, with the solid and dashed
lines indicating the mean oxygen abundances derived from the
infrared triplet $\lambda$7772-5 \AA\ and the oxygen abundance
obtained from the forbidden [O\,{\sc i}] $\lambda$6300 \AA\ line.
Thick lines in each of the panels correspond to  NLTE while thin
lines to LTE. The standard deviations ($\sigma$) for the observed
equivalent widths of the O\,{\sc i} and [O\,{\sc i}] lines are
shown in the middle panel. It is also of interest to note that our
LTE+1D calculations show that the lower the effective temperature
the larger the discrepancy in the iron abundance derived from Fe
{\sc i} and Fe {\sc ii} lines (see the thin lines in the middle
panel to the {\em lhs} of the figure).
\label{fig3}}
\end{figure*}

\clearpage

\begin{deluxetable}{lrrrr}
\tablecolumns{5} \tablewidth{0pc}
\tablecaption{O\,{\sc i}, Fe\,{\sc i} and Fe\,{\sc ii} line list and observed
equivalent widths ($W$) from the metal-poor star HD140283.
The oscillator strengths for the O\,{\sc i} IR triplet are from
Kurucz \& Bell (1995), while that for the forbidden [O\,{\sc i}]
line is from Lambert (1978).
The wavelengths and excitation energies of the Fe lines are from
Moore (1959).
The Fe oscillator strengths were taken from Fuhr et al. (1988),
except for Fe\,{\sc i} $\lambda$4733.596 \AA\ and Fe\,{\sc ii}
$\lambda\lambda$5197.569, 5234.620, 6149.238, 6456.376 \AA\ which
stem from Gurtovenko \& Kostik (1989). The observed equivalent widths
of all the Fe {\sc i} lines and of 11 Fe {\sc ii} lines have been kindly provided
to us by R. Garc\'\i a L\'opez (private communication), while the equivalent
widths of the remaining Fe {\sc ii} lines have been taken from Table 3 of
Nissen et al. (2002).}
\tablehead{
\colhead{Ion} & \colhead{$\lambda$ (\AA)}   & \colhead{log\, gf} &
\colhead{$\chi$ (eV)}& \colhead{$W$ (m\AA)}}
\startdata O\,{\sc i} & 6300.230 & $-9.750$ & 0.00 & 0.5\\
O\,{\sc i} & 7771.960 & $0.324$ & 9.11 & 7.9 \\
O\,{\sc i} & 7774.180 & $0.174$ & 9.11 & 4.8 \\
O\,{\sc i} & 7775.400 & $-0.046$ & 9.11 & 3.4\\
%
Fe\,{\sc i} & 3906.482 & $-2.243$ & 0.11 & 78.5\\
Fe\,{\sc i} & 3917.185 &$-2.155$  & 0.99 & 44.1 \\
Fe\,{\sc i} & 4005.246 & $-0.610$ & 1.55 & 79.4  \\
Fe\,{\sc i} & 4147.673 & $-2.104$ & 1.48 & 24.1\\
Fe\,{\sc i} & 4152.172 & $-3.232$ & 0.95 & 9.1  \\
Fe\,{\sc i} & 4177.597 & $-3.058$ & 0.91 & 24.3  \\
Fe\,{\sc i} & 4202.031 & $-0.708$ & 1.38 & 78.3 \\
Fe\,{\sc i} & 4216.186 & $-3.356$ & 0.00 & 38.9 \\
Fe\,{\sc i} & 4222.219 & $-0.967$ & 2.44 & 30.5 \\
Fe\,{\sc i} & 4250.125 & $-0.405$ & 2.46 & 52.4 \\
Fe\,{\sc i} & 4271.159 & $-0.349$ & 2.44 & 60.0 \\
Fe\,{\sc i} & 4375.932 & $-3.031$  & 0.00 & 55.0  \\
Fe\,{\sc i} & 4447.722 & $-1.342$  & 2.21 & 25.0 \\
Fe\,{\sc i} & 4489.741 & $-3.966$  & 0.12 & 11.2  \\
Fe\,{\sc i} & 4494.568 & $-1.136$ & 2.19 & 34.8  \\
Fe\,{\sc i} & 4733.596 & $-2.970$  & 1.48 & 5.5 \\
Fe\,{\sc i} & 4939.690 & $-3.340$ & 0.86 & 9.9 \\
Fe\,{\sc i} & 4994.133 & $-3.080$ & 0.91 & 13.8 \\
Fe\,{\sc i} & 5012.071 & $-2.642$  & 0.86 & 32.3 \\
Fe\,{\sc i} & 5083.342 & $-2.958$  & 0.95 & 15.5 \\
Fe\,{\sc i} & 5110.414 & $-3.760$  & 0.00 & 24.8 \\
Fe\,{\sc i} & 5123.723 & $-3.068$  & 1.01 & 12.3 \\
Fe\,{\sc i} & 5194.943 & $-2.090$ & 1.55 & 25.7 \\
Fe\,{\sc i} & 5198.714 & $-2.135$  & 2.21 & 8.5 \\
Fe\,{\sc i} & 5429.699 & $-1.879$ & 0.95 & 67.9 \\
Fe\,{\sc i} & 5701.553 & $-2.216$  & 2.55 & 2.5 \\
Fe\,{\sc i} & 6065.487 & $-1.530$  & 2.60 & 9.9 \\
Fe\,{\sc i} & 6136.620 & $-1.400$ & 2.44 & 19.7 \\
Fe\,{\sc i} & 6219.290 & $-2.433$  & 2.19 & 3.2 \\
Fe\,{\sc i} & 6252.561 & $-1.687$  & 2.39 & 14.4 \\
Fe\,{\sc i} & 6265.140 & $-2.550$ & 2.17 & 5.0 \\
Fe\,{\sc i} & 6593.878 & $-2.422$  & 2.42 & 5.5 \\
Fe\,{\sc i} & 6750.152 & $-2.621$ & 2.41 & 3.1 \\
Fe\,{\sc ii} & 4173.450 & $-2.180$ & 2.57 & 27.1  \\
Fe\,{\sc ii} & 4178.855 & $-2.480$ & 2.57 & 20.3 \\
Fe\,{\sc ii} & 4303.166 & $-2.490$ & 2.69 & 19.1 \\
Fe\,{\sc ii} & 4416.817 & $-2.600$ & 2.77 & 11.2 \\
Fe\,{\sc ii} & 4491.401 & $-2.700$  & 2.84 & 7.9 \\
Fe\,{\sc ii} & 4555.890 & $-2.290$ & 2.82 & 16.9 \\
Fe\,{\sc ii} & 4583.829 & $-2.020$ & 2.79 & 41.5 \\
Fe\,{\sc ii} & 4923.921 & $-1.320$ & 2.88 & 57.2 \\
Fe\,{\sc ii} & 5197.569 & $-2.380$ & 3.22 & 10.2 \\
Fe\,{\sc ii} & 5234.620 & $-2.310$ & 3.21 & 13.2 \\
Fe\,{\sc ii} & 5316.609 & $-1.850$ & 3.14 & 24.1 \\
Fe\,{\sc ii} & 6149.238 & $-2.850$ & 3.87 & 1.2   \\
Fe\,{\sc ii} & 6247.562 & $-2.510$ & 3.87 & 2.1  \\
Fe\,{\sc ii} & 6432.654 & $-3.740$  & 2.88 & 1.3  \\
Fe\,{\sc ii} & 6456.376 & $-2.280$ & 3.89 & 4.0 \\
%
\enddata
\end{deluxetable}


\begin{deluxetable}{lrccccc}
\tablecolumns{7}
\tablewidth{0pc}
\tablecaption{The solar oxygen and iron abundances obtained by
different authors using Asplund's et al. (2000) 
3D hydrodynamical model atmosphere and 
the 1D photospheric model
of Holweger \& M$\ddot{\rm u}$ller (1974).
The given uncertainties reflect only the line-to-line scatter.
The results are based on permitted
O\,{\sc i}\ IR lines, forbidden [O\,{\sc i}] lines, OH
vibration-rotation lines, OH pure rotation lines and lines of
Fe\,{\sc i}.}
\tablehead{
\colhead{Lines}   & \colhead{Approach}   &
\colhead{Present} &
\colhead{Asplund}   &
\colhead{Allende Prieto} &
\colhead{Shchukina \&} &
\colhead{Kostik}\\
\colhead{}   & &
\colhead{study} &
\colhead{et al.}   &
\colhead{et al.} &
\colhead{Trujillo Bueno} &
\colhead{et al.} \\
\colhead{}   & &
\colhead{} &
\colhead{(2004)}   &
\colhead{(2001)} &
\colhead{(2001)} & \colhead{(1996)}}
\startdata
%
 & & & & & & \\
 \multicolumn{7}{c}{3D solar model}
 \\
\hline
 & & & & & & \\
{O\,{\sc i} IR} & NLTE  & 8.70$\pm$0.06 & 8.65$\pm$0.01 & \nodata
& \nodata & \nodata
\\
{O\,{\sc i} IR}  & LTE & 8.93$\pm$0.06 & 8.89$\pm$0.02 & \nodata &
\nodata & \nodata
\\
{[O\,{\sc i}]} & LTE & \nodata & 8.68$\pm$0.01 & 8.69 &
\nodata & \nodata
\\
{OH vib-rot} & LTE  & \nodata & 8.61$\pm$0.03 & \nodata & \nodata
& \nodata
\\
{OH rot } & LTE     & \nodata & 8.65$\pm$0.02 &  \nodata & \nodata
& \nodata
\\
Fe\,{\sc i} & NLTE & \nodata  & \nodata  & \nodata & 7.50$\pm$0.10
& \nodata
\\
Fe\,{\sc i}& LTE & \nodata   & \nodata  & \nodata & 7.43$\pm$0.11
& \nodata
\\
\hline
 & & & & \\
 \multicolumn{7}{c}{1D model}\\
\hline
O\,{\sc i} & NLTE & 8.64$\pm$0.06 & 8.61$\pm$0.01  & \nodata & \nodata &
 \nodata
\\
Fe\,{\sc i} & LTE & \nodata  & \nodata  & \nodata & \nodata
& 7.62$\pm$0.04
\\
\enddata
\end{deluxetable}

\begin{deluxetable}{lrlll}
\tablecolumns{5}
\tablewidth{0pc}
\tablecaption{The oxygen and iron abundances for HD140283 obtained
by different authors using Asplund's et al. (1999) 
3D hydrodynamical model atmosphere and 1D hydrostatic
 models. 
The given uncertainties
reflect the line-to-line scatter. In general, the
abundances are based on different selection of lines, $gf$-values
and/or equivalent widths.
The results given in the table are based on
permitted
O\,{\sc i}\ IR lines, forbidden [O\,{\sc i}] lines, Fe\,{\sc i}
and  Fe\,{\sc ii} lines.}
\tablehead{
\colhead{Lines}   & \colhead{Approach}   &
\colhead{Present study} &
\colhead{Nissen et al. (2002)}   &
\colhead{Asplund et al. (1999)}}
 \startdata
%
  & & & & \\
 \multicolumn{5}{c}{3D model ($T_{\rm eff}=5700$ K, log\,g=3.7, [Fe/H]=$-$2.5)}
 \\
\hline
 & & & & \\
{O\,{\sc i} IR} & NLTE  & 7.08$\pm$0.06 & \nodata & \nodata
\\
{O\,{\sc i} IR}  & LTE  & 7.24$\pm$0.06 & \nodata & 7.20
($\lambda$7772)
\\
{[O\,{\sc i}]} & LTE    & 6.96          & 6.83    &  \nodata
\\
{Fe\,{\sc i}} & NLTE     & 5.77$\pm$0.19 & \nodata &  \nodata
\\
{Fe\,{\sc i}} & LTE     & 4.85$\pm$0.15 & \nodata  & 4.57$\pm$0.16
\\
{Fe\,{\sc ii}} & NLTE     & 5.60$\pm$0.48 & \nodata & \nodata
\\
{Fe\,{\sc ii}} & LTE     & 5.18$\pm$0.11 & 5.13$\pm$0.06 &
5.16$\pm$0.10
\\
\hline
 & & & & \\
 \multicolumn{5}{c}{1D models}\\
\hline
 & & $T_{\rm eff}=5700$  & $T_{\rm eff}=5690$  & $T_{\rm eff}=5700$
 \\
 & & log\,g=3.7 & log\,g=3.69 & log\,g=3.7 \\
 & & [Fe/H]=$-$2.5 & [Fe/H]=$-$2.42 & [Fe/H]=$-$2.5 \\
 \hline
   & & & & \\
{O\,{\sc i} IR} & NLTE  & 7.11$\pm$0.06 & 7.02 & \nodata
\\
{O\,{\sc i} IR}  & LTE  & 7.29$\pm$0.06 & 7.11 & 7.15
($\lambda$7772)
\\
{[O\,{\sc i}]} & LTE    & 7.18          & 7.09    &  \nodata
\\
{Fe\,{\sc i}} & NLTE     & 5.77$\pm$0.15 & \nodata &  \nodata
\\
{Fe\,{\sc i}} & LTE     & 5.20$\pm$0.14 & \nodata  & 5.02$\pm$0.17
\\
{Fe\,{\sc ii}} & NLTE     & 5.34$\pm$0.21 & \nodata & \nodata
\\
{Fe\,{\sc ii}} & LTE     & 5.18$\pm$0.11 & 5.08$\pm$0.06 &
5.08$\pm$0.11
\\
\enddata
\end{deluxetable}


\clearpage

\begin{deluxetable}{lrrr}
\tablecolumns{4} \tablewidth{0pc}
\tablecaption{Stellar parameters of the star HD140283}
\tablehead{
\colhead{Case} & \colhead{$\rm{T}_{eff}$ (K)}   & \colhead{[Fe/H]}
& \colhead{log$g$}}
 \startdata 1D+LTE  & 5670 & $-2.3$ & 3.7 \\
            1D+NLTE & 5450 & $-2.1$ & 3.7 \\
            3D+LTE  & 5700 & $-2.3$ & 3.7 \\
            3D+NLTE & 5600 & $-2.0$ & 3.7 \\
\enddata
\end{deluxetable}

\end{document}